\documentclass[twocolumn,english,amssymb,prl,superscriptaddress,notitlepage]{revtex4-2}
\usepackage{amsmath,amssymb,graphicx}
\usepackage{multirow}
\usepackage{fancyhdr,bm}
\usepackage{booktabs} 
\usepackage{braket}
\usepackage{titlesec}
\titleformat{\paragraph}
{\normalfont\normalsize\bfseries}{\theparagraph}{1em}{}
\titlespacing*{\paragraph}
{0pt}{3.25ex plus 1ex minus .2ex}{1.5ex plus .2ex}
 
\usepackage{color}
\usepackage[colorlinks=true, pdfstartview=FitV, linkcolor=blue, citecolor=blue, urlcolor=blue]{hyperref}

\begin{document}
\title{Isoperimetric Inequalities in Quantum Geometry}
\author{Praveen Pai}
\author{Fan Zhang}\email{zhang@utdallas.edu}
\affiliation{Department of Physics, University of Texas at Dallas, Richardson, Texas 75080, USA}
\date{\today}
\begin{abstract}
We reveal strong and weak inequalities relating two fundamental macroscopic quantum geometric quantities, the quantum distance and Berry phase, 
for closed paths in the Hilbert space of wavefunctions.
We recount the role of quantum geometry in various quantum problems and show that our findings place new bounds on important physical quantities.
\end{abstract}
\maketitle

\section{Introduction}

Many fundamental concepts in geometry hold the keys to understanding our nature and universe, 
from Einstein's general relativity of space and time to the standard model of elementary particles.
As for the light and matter in between, the physical properties in both ground and excited states 
are often shaped by the quantum geometric properties of particles' wavefunctions in Hilbert space, 
such as Berry curvature and quantum metric~\cite{provostRiemannianStructureManifolds1980}. Examples are abundant: 
various Hall effects~\cite{nagaosaAnomalous2010,sinovaSpin2015,gaoFieldInducedPositional2014,gaoQuantum2023}, flat-band superconductivity~\cite{peottaSuperfluidityTopologicallyNontrivial2015,rossiQuantum2021}, orbital magnetism~\cite{zhangSpontaneousQuantumHall2011}, 
resonant optical responses~\cite{ahnRiemannianGeometryResonant2022}, and not to mention topological phases of matter~\cite{hasanColloquium2010}.

In the same vein, we may seek to reveal deep implications of the isoperimetric problem. 
On the flat two-dimensional (2D) plane, 
what closed shape maximizes area for a fixed perimeter, and what is the resultant area to perimeter ratio? Though the solution is intuitively a circle, 
this surprisingly had been an outstanding problem in mathematics until it was rigorously proven in the 19th century~\cite{blasjoIsoperimetric2005a}. 
Since then, the isoperimetric problem has been generalized to hypervolumes and hypersurfaces 
in higher-dimensional Euclidean spaces, spheres, as well as more exotic manifolds~\cite{ossermanIsoperimetricInequality1978}. 
With elegant relationships between area and perimeter, 
the isoperimetric problem motivated the development of the calculus of variations, which helped us ascertain many classical principles of our nature and universe~\cite{gelfandCalculusVariations2000}. 

One may naturally wonder whether there exists an analog of the isoperimetric problem in quantum geometry.
If the answer is in the affirmative, what are its implications in quantum physics? 
Here we accomplish this by first mapping the two-band quantum isoperimetric problem directly to the spherical isoperimetric problem, 
revealing a strong inequality relating two fundamental macroscopic quantum geometric quantities, the quantum distance and the Berry phase. 
Furthermore, we demonstrate a weak inequality for the most general multi-band case: 
the quantum distance is no smaller than the Berry phase for any closed path in the Hilbert space of wavefunctions,
and that the strong inequality is not violated in this case for infinitesimal variations of curves known to saturate it. 
We conclude by applying our new isoperimetric inequalities to a variety of different quantum systems, 
placing new bounds on important physical quantities, including Wannier function spread, quantum speed limit, electron-phonon coupling, and geometric superfluid weight.

\begin{figure}[t]
\includegraphics[width=1\columnwidth]{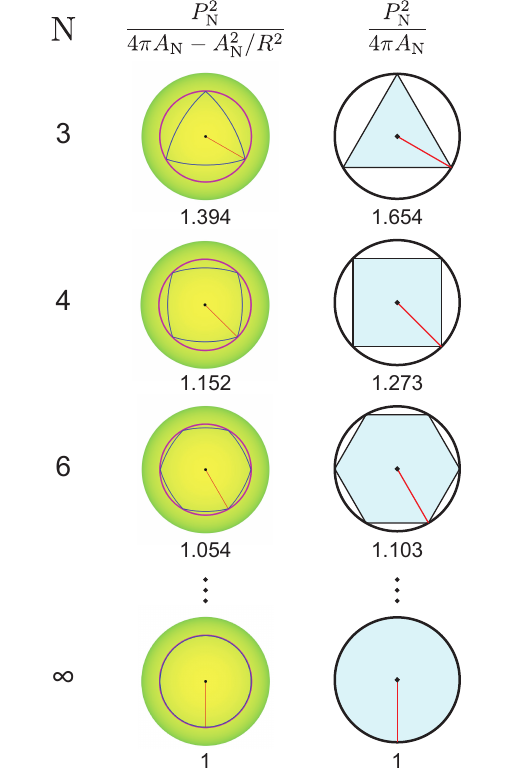}
\caption{Visual of the 2D isoperimetric inequalities. Right: equilateral triangle, square, hexagon, and circle of equal radius (red lines). 
As the number of sides, $N$, of a regular polygon increases, the inverse isoperimetric quotient (shown at the top) saturates to $1$, 
revealing that the planar isoperimetric problem is satisfied by the circle. 
Left: the same visual but for spherical polygons on a sphere with their radial lengths fixed at the polar angle of ${\pi}/{4}$. 
The spherical isoperimetric problem is again solved by a shape with the highest symmetry -- a closed path of constant polar angle.} 
\label{fig1}
\end{figure}

\section{Classical Isoperimetric Inequalities}

We now recount the isoperimetric problem in classical geometry. Proven in 1879~\cite{blasjoIsoperimetric2005a}, 
the original isoperimetric problem on the 2D plane ($\mathbb{R}^2$) revealed a universal inequality 
between the area $A$ and perimeter $P$ of any closed loop in the plane:
\begin{equation}
P^2 \geq 4\pi A\,.\label{plane}
\end{equation}
The shape that saturates this inequality is circle, as can be verified by substituting $P =2\pi r$ and $A = \pi r^2$, 
where $r$ is the circle's radius. Equation~(\ref{plane}) is known as the isoperimetric inequality for the plane. 
A regular polygon with $N$ sides perfectly illustrates this inequality, as shown in Fig.~\ref{fig1}, where 
the inverse isoperimetric quotient reads $P^2/(4\pi A) = (N/\pi) \tan(\pi/N)$, 
amounting to $1.7,~1.3,~1.1$,~and~$1$ respectively for equilateral triangle, square, hexagon, and circle (the infinite $N$ case).

Historically, it was natural to pose the same isoperimetric problem on curved spaces such as the 2D sphere ($S^2$). 
However, the isoperimetric problem on the 2D sphere was not solved until 1905~\cite{ossermanIsoperimetricInequality1978}, 
and the generalized inequality reads
\begin{equation}
P^2 \geq 4\pi A-\frac{A^2}{R^2}\,,\label{sphere}
\end{equation}
where $A$ is the surface area enclosed by a loop of perimeter $P$ on the sphere, and $R$ is the radius of the sphere. 
The shape that saturates the isoperimetric inequality is again a ``circle'', which can be identified as a closed path of constant polar angle on the sphere. 
This inequality can be intuitively divined from the limit of spherical polygons with their vertices sharing the same polar angle $\theta$, 
as illustrated in Fig.~\ref{fig1} for the polar angle of $\theta=45^\circ$ and detailed in Supplementary Materials (SM).  

\section{Quantum Isoperimetric Inequalities}

In the quantum regime, we start from a familiar definition of the gauge-invariant quantum geometric tensor~\cite{provostRiemannianStructureManifolds1980} commonly used in physics:
\begin{equation}
    \chi_{\mu \nu} = \langle{\partial_\mu \psi}|{\partial_\nu \psi}\rangle - \langle{\partial_\mu \psi}|{ \psi}\rangle \langle{\psi}|{\partial_\nu \psi}\rangle\,,\label{qgt}
\end{equation}
which is the Fubini-Study metric evaluated at an arbitrary pure state $\ket{\psi}$. In terms of real and imaginary components, $\chi_{\mu \nu} = g_{\mu \nu}-\frac{i}{2}F_{\mu \nu}$, where $F_{\mu \nu}$ is the antisymmetric Berry curvature, and $g_{\mu \nu}$ is the symmetric quantum metric \cite{chengQuantum2010,onishiFundamentalBoundTopological2024}.
The Greek indices label the coordinates by which the Hilbert space is parametrized. 
Note that Eq.~(\ref{qgt}) is not necessarily a momentum-space parameterization, 
as we seek to identify universal properties of Hilbert space,
although the use of crystal momentum has facilitated many recent developments~\cite{ahnRiemannianGeometryResonant2022,peottaSuperfluidityTopologicallyNontrivial2015,yuNontrivialQuantumGeometry2024,onishiFundamentalBoundTopological2024}.
For a single-band system, the Hilbert space is a trivial point and $\chi=0$. 

\begin{figure*}[t]
\centering
\includegraphics[scale=1]{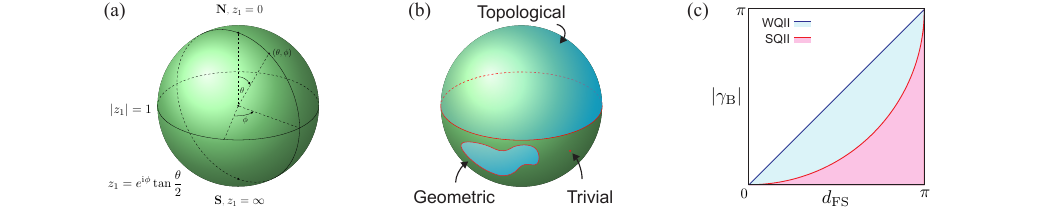}
\caption{Illustration of Bloch sphere parameterization and quantum isoperimetric inequalities. 
(a) Any two-band state can be expressed as $\ket{z_1} = (1,z_1)^{T}/\sqrt{1+|{z_1}|^2}$ 
using one complex number $z_1=e^{i\phi}\tan(\theta/2)$, 
where $\phi$ is the azimuthal angle and $\theta$ is the polar angle of the Bloch sphere. 
$\theta=0$ ($z_1=0$) and $\pi$ ($z_1=\infty$) denote the the north and south poles of the Bloch sphere, respectively. 
(b) Representatives of trivial, geometric, and topological loops. 
(c) Geometric implication of the weak quantum isoperimetric inequality derived from the strong one.}
\label{fig2}

\end{figure*}

For a two-band system, a pure state can be represented by two complex numbers and is well-defined up to normalization and a phase factor.
Thus, the relevant Hilbert space is the complex projective space $\mathbb{C}P^1$~\cite{nakaharaGeometryTopologyPhysics2005},
topologically equivalent to $S^2$ and known as the Bloch sphere. 
The two special states $(1,0)^{T}$ and $(0,1)^{T}$ can be defined as the north and south poles of the Bloch sphere, respectively.
It follows that any state can be expressed as $\ket{z_1} = (1,z_1)^{T}/\sqrt{1+|{z_1}|^2}$ 
using one complex number $z_1=e^{i\phi}\tan(\theta/2)$, 
where $\phi$ is the azimuthal angle and $\theta$ is the polar angle of the Bloch sphere, 
with $\theta=0$ and $\pi$ for the two poles, as illustrated in Fig.~\ref{fig2}a. 
Correspondingly, the quantum geometric tensor is just the Fubini-Study metric $\chi_{11} = 1/(1+|{z_1}|^2)^{2}$. 
This is equivalent to the 3D Euclidean metric projected onto the 2D sphere of radius ${1}/{2}$, i.e., the Bloch sphere. 
It is straightforward to show that the geodesic distance between the north and south poles is $\pi/2$, i.e., one half of the perimeter of great circles.
More generally, the geodesic distance between the north pole and the state $| z_1 \rangle$ on the Bloch sphere reads 
$\arccos(|\langle 0 | z_1 \rangle|) = \theta/2\leq\pi/2$, which provides a lower bound for the quantum distance $d_{\rm FS}$ 
calculated through the Fubini-Study metric, $ds^2 = \chi_{11}dz_1d\Bar{z}_1$, for any arbitrary path between them.

With this metric, the Bloch sphere admits a neat relationship between the Berry phase $\gamma_{\rm B}$ 
of a loop on the sphere and its enclosed solid angle $\Omega$: $|\gamma_{\rm B}|= {\Omega}/{2}$~\cite{xiaoBerryPhaseEffects2010}. Due to the compact nature of the Bloch sphere, there is ambiguity in the choice of solid angle enclosed by the loop. 
This is related to the fact that Berry phase is only gauge invariant mod $2\pi$~\cite{xiaoBerryPhaseEffects2010}. 
Hereafter,  we choose $\gamma_{\rm B} \in (-\pi,\pi]$, corresponding to the smaller of the two possible solid angles. 
(One may naively note that since the metric $\chi_{11}$ is real, there is no Berry curvature distributed over the Bloch sphere. 
However, this neglects the fact that if the metric is written in terms of $\theta$ and $\phi$, it does have an imaginary part--Berry curvature.)

As the Fubini-Study metric identifies the Bloch sphere as the Euclidean sphere of radius ${1}/{2}$,
substituting $A =\Omega R^2$, $P = d_{\rm FS}$, and $R={1}/{2}$ in Eq.~\eqref{sphere} yields the following quantum isoperimetric inequality: 
\begin{equation}
    (|\gamma_{\rm B}|-\pi)^2 +d_{\rm FS}^2 \geq \pi^2\,.\label{strong}
\end{equation}
The equality occurs for all the circles on the Bloch sphere.
As shown in Fig.~\ref{fig2}c, any $(d_{\rm FS}, |\gamma_{\rm B}|)$ corresponding to a loop on the Bloch sphere 
falls beneath the quarter circle specified by the equality of Eq.~(\ref{strong}). 
Because the chord $d_{\rm FS} = |\gamma_{\rm B}|$ connecting the same two ends is above the quarter circle, 
all such $(d_{\rm FS}, |\gamma_{\rm B}|)$ also satisfy a weaker inequality:
\begin{equation}
d_{\rm FS}\geq\gamma_{\rm B}\,.\label{weak}
\end{equation}
The equality only occurs for $d_{\rm FS}=\gamma_{\rm B}=0$ or $d_{\rm FS}=\gamma_{\rm B}=\pi$. 
A great circle of the Bloch sphere exemplifies the latter case, 
which can be realized in 1D systems such as the Su–Schrieffer–Heeger (SSH) model 
and 1D subsystems such as the monolayer graphene
with a nontrivial topological winding number $\pm 1$ protected by chiral symmetry~\cite{zhangSpontaneousQuantumHall2011}. 
Hereafter we shall dub Eqs.~(\ref{strong}) and~(\ref{weak}) the strong and weak quantum isoperimetric inequalities (QII), respectively. 
The weak QII is yet another example of inequalities in condensed matter physics that yields 
equality between geometrical and topological quantities in the presence of a special symmetry~\cite{yuNontrivialQuantumGeometry2024,onishiFundamentalBoundTopological2024,peottaSuperfluidityTopologicallyNontrivial2015}. However, the equality of the strong QII may occur for non topological cases in the absence of chiral symmetry.

The weak QII can readily be extended to loops that self-intersect. 
If a loop intersects with itself at a point, it may be broken into two sub-loops, each with $\gamma_{\rm B} \in (-\pi,\pi]$. 
In the case of many self-intersections, the weak QII Eq.~(\ref{weak}) applies to each sub-loop as labeled by $i$, $d_{\rm FS}^{(i)}\geq \gamma_{\rm B}^{(i)}$. 
As such, the accumulated Berry phase is bounded above by the total quantum distance, $\sum_i d_{\rm FS}^{(i)}\geq \sum_i \gamma_{\rm B}^{(i)}$.
A prime example of this occurring is in $N$-layer rhombohedral graphene~\cite{zhangSpontaneousQuantumHall2011}. 
When an electron traverses around a single Dirac point once in momentum space, because of chiral symmetry, 
its wavefunction travels around the equator of the Bloch sphere $N$ times, yielding a winding number of $N$ and accumulatively $d_{\rm FS} = \gamma_{\rm B} = N\pi$.

For $M$-band systems, the Hilbert space can be represented by the complex projective space $\mathbb{C}P^{M-1}$~\cite{nakaharaGeometryTopologyPhysics2005}, 
and the states can be parameterized by $M-1$ complex variables $\bm{z}=(z_1,z_2,...z_{M-1})$, 
as natural generalizations of the Bloch sphere and $z_1$ parameterization (Fig.~\ref{fig2}a) for two-band systems. 
Significantly, the weak QII Eq.~(\ref{weak}) is valid for $M\geq2$, as demonstrated in SM with a different approach.
Similarly in SM, the equality of Eq.~(\ref{strong}) is also shown to hold for circles in $\mathbb{C}P^{\rm M-1}$, 
since $\mathbb{C}P^1$ is a subset of $\mathbb{C}P^{\rm M-1}$. Furthermore, circles are shown to be extremal for Berry phases, 
as any infinitesimal deviation of a circle does not change the corresponding Berry phase to first order. 
As such, we conjecture that the strong QII Eq.~(\ref{strong}) also holds for $M>2$. In the rest of work, 
we seek to demonstrate the power of the weak QII in placing new bounds on physical quantities. 

\section{Applications}

\subsection{Wannier Function Spread} 

Quantum distance first appeared in condensed matter physics in the context of Wannier function spread
in the landmark work by Marzari and Vanderbilt~\cite{marzariMaximallyLocalizedGeneralized1997} on maximally localized Wannier functions.
Wannier functions provide a valuable tool in theoretical and computational physics, 
leading to the modern theory of polarization~\cite{king-smithTheoryPolarizationCrystalline1993} and many efficient computational techniques.
The Wannier function spread denotes the standard deviation of the Wannier functions associated with a particular band. 
For this reason, we choose to examine how quantum geometry bounds the Wannier function spread.

The Wannier function spread can be separated into two parts: 
one that is gauge invariant and one that is not but vanishes for 1D systems~\cite{marzariMaximallyLocalizedGeneralized1997}.  
The gauge-invariant part reads
\begin{equation}
    \Omega_1 = \frac{\omega}{(2\pi)^d}\int {\rm Tr}[g(\bm{k})]d^d{\bm k}\,,
\end{equation}
where $\omega$ is the volume of unit cell of dimension $d$. 
Using the Cauchy-Schwartz inequality (see SM) and the weak QII, 
it is straightforward to show for 1D systems
\begin{equation}
    \Omega_1 \geq (\frac{a\,d_{\rm FS}}{2\pi})^2 \geq (\frac{a\,\gamma_{\rm B}}{2\pi})^2\,,
    \label{wannier}
\end{equation}
where $a$ is the lattice constant. This result is especially interesting because it implies that the Wannier function spread 
is lower bounded by both the quantum distance squared and the Berry phases squared. 
The latter naturally limits the maximal localization of Wannier functions by polarization, 
whereas the former provides a better bound, given the weak QII. 
Note that our result can be generalized to higher dimensions (see SM).

\subsection{Quantum Speed Limit}

The Heisenberg uncertainty principle is a cornerstone of quantum mechanics. 
It states that there is a limit to the precision with which certain pairs of physical properties, 
such as position and momentum, can be simultaneously known. 
However, the energy–time uncertainty relation $\Delta E \Delta t \geq {\hbar}/{2}$ has a long, controversial history~\cite{deffnerQuantumSpeedLimits2017}. With the experimentally verified connection between the life-time and energy width of a resonance state~\cite{bohmRelativistic2005},
the accepted interpretation of this relation leads to the so-called quantum speed limit~\cite{mandelstamUncertainty1991,margolusmaximum1998,deffnerQuantumSpeedLimits2017} 
on the time that it takes for an initial state to evolve into a final state. 
This fundamental limit reveals how quickly a coherent state may evolve, presenting building guidance and a performance limit for quantum computers. 

Usually quantum speed limits are written for evolution between orthogonal states. 
Since quantum geometry considers not only the destination but also the journey in the evolution, 
we now reveal a new, geometric phase limit. Anandan and Aharonov found~\cite{anandanGeometryQuantumEvolution1990} 
the following relation between quantum distance and energy uncertainty:
\begin{equation}
    \frac{d d_{\rm FS}}{dt} = \frac{\Delta E}{\hbar}\,,\label{aa}
\end{equation}
where $(\Delta E)^2 = \langle H(t)^2\rangle-\langle H(t) \rangle ^2$ for Hamiltonian $H$ at time $t$. 
In general, the initial state is not an eigenstate of the initial Hamiltonian, 
in sharp contrast to the case of Bloch states over a Brillouin zone. 
By integrating Eq.~(\ref{aa}) and then applying the weak QII, 
we obtain the following quantum speed limit imposed by the Berry phase of a cyclic adiabatic evolution:
\begin{equation}
    \tau \geq \frac{\gamma_{\rm B} \hbar}{\langle \Delta E \rangle}\,,
\end{equation}
with $\langle \Delta E \rangle = \int_0^\tau \Delta E dt /{\tau} $.
This geometric phase limit provides the exact quantum speed when the equality occurs for the weak QII.

\subsection{Electron-Phonon Coupling} 

In phonon-mediated superconductors the transition temperature $T_c$ is related to 
the electron-phonon coupling constant $\lambda$~\cite{allenTransitionTemperatureStrongcoupled1975,nambuQuasiParticlesGaugeInvariance1960}. 
As $T_c$ generally increases with $\lambda$, identifying upper and lower bounds of $\lambda$ is important for maximizing $T_c$.   
Recently, it has been shown~\cite{yuNontrivialQuantumGeometry2024} that the quantum geometry of the Fermi surface can contribute significantly to $\lambda$, and two prime examples are 2D graphene and 3D MgB$_2$.
For 2D systems Fermi surfaces are generally 1D. Using the simplest model -- the {gapless} Dirac model --  
we shall show $\lambda_{\rm geo}$, the quantum geometric contribution to $\lambda$, are related to $d_{\rm FS}$ and $\gamma$. 
For this model~\cite{yuNontrivialQuantumGeometry2024}, 
\begin{equation}
     \lambda_{\rm geo} \propto \frac{E_F^2}{v_F}\int_{\rm FS}  {\rm Tr}[g({\bm k})]d\sigma_{\bm k}\,,
\end{equation}
where $E_F$ and $v_F$ are Fermi energy and Fermi velocity. 
At any point of the Fermi surface, the metric can be orthogonally transformed 
so that the first component corresponds to the one along the Fermi surface, $k_{\ell}$. 
As such, ${\rm Tr}[g({\bm k})]$ is lower bounded by $g_{\ell\ell}$, 
since the metric diagonal elements are all nonnegative.
Following the same argument as Eq.~(\ref{wannier}), 
\begin{equation}
     \int_{\rm FS}  \!{\rm Tr}[g({\bm k})]d\sigma_{\bm k} \geq \int\!\! g_{\ell\ell}({\bm k})dk_\ell 
     \geq  \ell_{\rm FS}^{-1} d_{\rm FS}^2 \geq  \ell_{\rm FS}^{-1} \gamma_{\rm B}^2\,,\label{FS}
\end{equation}
where $\ell_{\rm FS}$ is the Fermi-surface perimeter.
Evidently, $d_{\rm FS}^2$ provides a more direct and better bound than $\gamma_{\rm B}^2$.

For the gapless case such as graphene, the Berry curvature vanishes, yet the quantum metric reads
\begin{equation}
    g({\bm k}) = \frac{1}{4k^4}\begin{bmatrix}
        {k_y^2} & -{k_x k_y} \\
        -{k_x k_y} & {k_x^2}
    \end{bmatrix}\,,
\end{equation}
yielding ${\rm Tr}[g(\bm{k})]=g_{\ell\ell}={1}/{4k^2}$ and $g_{rr}=0$. 
As $d_{\rm FS} = \gamma_{\rm B} = \pi$ and $\ell_{\rm FS}=2\pi E_F/v_F$, 
Eq.~(\ref{FS}) exhibits three equalities and reaches the topological bound $\pi v_F/2E_F$.

\subsection{Geometric Superfluid Weight}

Recently, it has been discovered that the quantum metric can contribute to the superfluid weight $D_s$. 
A prime example is an isolated 2D flat band, where $D_s$ is bounded below by its Chern number~\cite{peottaSuperfluidityTopologicallyNontrivial2015}, 
despite the vanishing intra-band contribution due to the flatness.
In particular, $D_s$ for an attractive Hubbard model is given 
by the integral of ${\rm Tr}[g({\bm k})]$ over the Brillouin zone in any dimension. 

For 1D systems~\cite{tovmasyanEffectiveTheoryEmergent2016}, 
\begin{equation}
    D_s = \frac{U}{\pi^2\hbar^2 M}\nu(1-\nu) \int {\rm Tr}[g({k})] dk\,, \label{sfweight}
\end{equation}
where $U$ is the interaction strength, $M$ is the number of bands, and $\nu$ is the filling factor. 
Similar to the previous examples, it is straightforward to show 
\begin{equation}
    D_s \geq \frac{aU}{2\pi^3\hbar^2 M}\nu(1-\nu) d_{\rm FS}^2 \geq \frac{aU}{2\pi^3\hbar^2 M}\nu(1-\nu) \gamma_{\rm B}^2. \label{sfweightineq}
\end{equation} 
While the topological winding number bound dictating $d_{\rm FS} = \gamma_{\rm B}$ 
was derived before for a specific two-band model with chiral symmetry~\cite{tovmasyanEffectiveTheoryEmergent2016},  
our geometric phase bound is more general and valid for 1D systems with any number of bands regardless of any symmetry. 
Moreover, quantum distance provides a better bound, given the weak QII. 
Importantly, the problem of maximizing $\gamma_{\rm B}$, a gauge-dependent quantity, to potentially increase $D_s$ 
becomes a matter of maximizing $d_{\rm FS}$, a gauge-independent quantity.

A recent effort shows that the superfluid weight is only related to the minimal quantum metric that minimizes $\Omega_1$. 
In the SSH model~\cite{suSolitonsPolyacetylene1979}, the two extreme limits of dimerization yields flat bands, 
with $\gamma_{\rm B}$ quantized to $0$ and $\pi$ relatively. 
Physically, the two cases are identical in that they are related by a half unit-cell shift. 
Thus, we can conclude that $d_{\rm FS} = \gamma_{\rm B} = 0$ in Eq.~(\ref{sfweightineq}), since it is the minimal quantum metric that plays a physical role.
The original Creutz ladder~\cite{creutzEndStatesLadder1999} is another 1D model 
that features a $\pi$ magnetic flux per unit cell and two topological flat bands. 
Unlike the SSH model, the minimal quantum metric produces $d_{\rm FS} =\gamma_{\rm B} = \pi$.
Indeed, this nonzero $D_s$ has been found by both analytical~\cite{tovmasyanEffectiveTheoryEmergent2016} and numerical~\cite{mondainiPairingSuperconductivityFlat2018} calculations. 
\section{Conclusion}
Through our strong and weak isoperimetric inequalities, we have identified relationships between quantum distance and Berry phase. Notably, these results hold true even without assuming symmetries, highlighting a general aspect of the differentiable structure of wavefunctions. Despite a century of quantum mechanics and decades since the introduction of the quantum geometric tensor~\cite{provostRiemannianStructureManifolds1980}, these findings might offer a fresh perspective. Quantum geometry continues to be an active area of research in condensed matter physics due to its relevance to the structure of many-body phases. As a result, we are encouraged to revisit fundamental concepts—such as the Berry phase and quantum distance—and thoughtfully reconsider basic aspects of quantum theory, for there may be elegant relations left unwritten.

\bibliographystyle{apsrev4-2_edit_200521}
\bibliography{refs}

\begin{thebibliography}{33}%
\makeatletter
\providecommand \@ifxundefined [1]{%
 \@ifx{#1\undefined}
}%
\providecommand \@ifnum [1]{%
 \ifnum #1\expandafter \@firstoftwo
 \else \expandafter \@secondoftwo
 \fi
}%
\providecommand \@ifx [1]{%
 \ifx #1\expandafter \@firstoftwo
 \else \expandafter \@secondoftwo
 \fi
}%
\providecommand \natexlab [1]{#1}%
\providecommand \textit  [1]{``#1''}%
\providecommand \bibnamefont  [1]{#1}%
\providecommand \bibfnamefont [1]{#1}%
\providecommand \citenamefont [1]{#1}%
\providecommand \href@noop [0]{\@secondoftwo}%
\providecommand \href [0]{\begingroup \@sanitize@url \@href}%
\providecommand \@href[1]{\@@startlink{#1}\@@href}%
\providecommand \@@href[1]{\endgroup#1\@@endlink}%
\providecommand \@sanitize@url [0]{\catcode `\\12\catcode `\$12\catcode `\&12\catcode `\#12\catcode `\^12\catcode `\_12\catcode `\%12\relax}%
\providecommand \@@startlink[1]{}%
\providecommand \@@endlink[0]{}%
\providecommand \url  [0]{\begingroup\@sanitize@url \@url }%
\providecommand \@url [1]{\endgroup\@href {#1}{\urlprefix }}%
\providecommand \urlprefix  [0]{URL }%
\providecommand \Eprint [0]{\href }%
\providecommand \doibase [0]{https://doi.org/}%
\providecommand \selectlanguage [0]{\@gobble}%
\providecommand \bibinfo  [0]{\@secondoftwo}%
\providecommand \bibfield  [0]{\@secondoftwo}%
\providecommand \translation [1]{[#1]}%
\providecommand \BibitemOpen [0]{}%
\providecommand \bibitemStop [0]{}%
\providecommand \bibitemNoStop [0]{.\EOS\space}%
\providecommand \EOS [0]{\spacefactor3000\relax}%
\providecommand \BibitemShut  [1]{\csname bibitem#1\endcsname}%
\let\auto@bib@innerbib\@empty
\bibitem [{\citenamefont {Provost}\ and\ \citenamefont {Vallee}(1980)}]{provostRiemannianStructureManifolds1980}%
  \BibitemOpen
  \bibfield  {author} {\bibinfo {author} {\bibfnamefont {J.~P.}\ \bibnamefont {Provost}}\ and\ \bibinfo {author} {\bibfnamefont {G.}~\bibnamefont {Vallee}},\ }\bibfield  {title} {\textit {\bibinfo {title} {Riemannian Structure on Manifolds of Quantum States},}\ }\href {https://doi.org/10.1007/BF02193559} {\bibfield  {journal} {\bibinfo  {journal} {Commun. Math. Phys.}\ }\textbf {\bibinfo {volume} {76}},\ \bibinfo {pages} {289--301} (\bibinfo {year} {1980})}\BibitemShut {NoStop}%
\bibitem [{\citenamefont {Nagaosa}\ \emph {et~al.}(2010)\citenamefont {Nagaosa}, \citenamefont {Sinova}, \citenamefont {Onoda}, \citenamefont {MacDonald},\ and\ \citenamefont {Ong}}]{nagaosaAnomalous2010}%
  \BibitemOpen
  \bibfield  {author} {\bibinfo {author} {\bibfnamefont {N.}~\bibnamefont {Nagaosa}}, \bibinfo {author} {\bibfnamefont {J.}~\bibnamefont {Sinova}}, \bibinfo {author} {\bibfnamefont {S.}~\bibnamefont {Onoda}}, \bibinfo {author} {\bibfnamefont {A.~H.}\ \bibnamefont {MacDonald}},\ and\ \bibinfo {author} {\bibfnamefont {N.~P.}\ \bibnamefont {Ong}},\ }\bibfield  {title} {\textit {\bibinfo {title} {Anomalous {{Hall}} Effect},}\ }\href {https://doi.org/10.1103/RevModPhys.82.1539} {\bibfield  {journal} {\bibinfo  {journal} {Rev. Mod. Phys.}\ }\textbf {\bibinfo {volume} {82}},\ \bibinfo {pages} {1539--1592} (\bibinfo {year} {2010})}\BibitemShut {NoStop}%
\bibitem [{\citenamefont {Sinova}\ \emph {et~al.}(2015)\citenamefont {Sinova}, \citenamefont {Valenzuela}, \citenamefont {Wunderlich}, \citenamefont {Back},\ and\ \citenamefont {Jungwirth}}]{sinovaSpin2015}%
  \BibitemOpen
  \bibfield  {author} {\bibinfo {author} {\bibfnamefont {J.}~\bibnamefont {Sinova}}, \bibinfo {author} {\bibfnamefont {S.~O.}\ \bibnamefont {Valenzuela}}, \bibinfo {author} {\bibfnamefont {J.}~\bibnamefont {Wunderlich}}, \bibinfo {author} {\bibfnamefont {C.~H.}\ \bibnamefont {Back}},\ and\ \bibinfo {author} {\bibfnamefont {T.}~\bibnamefont {Jungwirth}},\ }\bibfield  {title} {\textit {\bibinfo {title} {Spin {{Hall}} Effects},}\ }\href {https://doi.org/10.1103/RevModPhys.87.1213} {\bibfield  {journal} {\bibinfo  {journal} {Rev. Mod. Phys.}\ }\textbf {\bibinfo {volume} {87}},\ \bibinfo {pages} {1213--1260} (\bibinfo {year} {2015})}\BibitemShut {NoStop}%
\bibitem [{\citenamefont {Gao}\ \emph {et~al.}(2014)\citenamefont {Gao}, \citenamefont {Yang},\ and\ \citenamefont {Niu}}]{gaoFieldInducedPositional2014}%
  \BibitemOpen
  \bibfield  {author} {\bibinfo {author} {\bibfnamefont {Y.}~\bibnamefont {Gao}}, \bibinfo {author} {\bibfnamefont {S.~A.}\ \bibnamefont {Yang}},\ and\ \bibinfo {author} {\bibfnamefont {Q.}~\bibnamefont {Niu}},\ }\bibfield  {title} {\textit {\bibinfo {title} {Field {{Induced Positional Shift}} of {{Bloch Electrons}} and {{Its Dynamical Implications}}},}\ }\href {https://doi.org/10.1103/PhysRevLett.112.166601} {\bibfield  {journal} {\bibinfo  {journal} {Phys. Rev. Lett.}\ }\textbf {\bibinfo {volume} {112}},\ \bibinfo {pages} {166601} (\bibinfo {year} {2014})}\BibitemShut {NoStop}%
\bibitem [{\citenamefont {Gao}\ \emph {et~al.}(2023)\citenamefont {Gao}, \citenamefont {Liu}, \citenamefont {Qiu}, \citenamefont {Ghosh}, \citenamefont {V.~Trevisan}, \citenamefont {Onishi}, \citenamefont {Hu}, \citenamefont {Qian}, \citenamefont {Tien}, \citenamefont {Chen}, \citenamefont {Huang}, \citenamefont {B{\'e}rub{\'e}}, \citenamefont {Li}, \citenamefont {Tzschaschel}, \citenamefont {Dinh}, \citenamefont {Sun}, \citenamefont {Ho}, \citenamefont {Lien}, \citenamefont {Singh}, \citenamefont {Watanabe}, \citenamefont {Taniguchi}, \citenamefont {Bell}, \citenamefont {Lin}, \citenamefont {Chang}, \citenamefont {Du}, \citenamefont {Bansil}, \citenamefont {Fu}, \citenamefont {Ni}, \citenamefont {Orth}, \citenamefont {Ma},\ and\ \citenamefont {Xu}}]{gaoQuantum2023}%
  \BibitemOpen
  \bibfield  {author} {\bibinfo {author} {\bibfnamefont {A.}~\bibnamefont {Gao}}, \bibinfo {author} {\bibfnamefont {Y.-F.}\ \bibnamefont {Liu}}, \bibinfo {author} {\bibfnamefont {J.-X.}\ \bibnamefont {Qiu}}, \bibinfo {author} {\bibfnamefont {B.}~\bibnamefont {Ghosh}}, \bibinfo {author} {\bibfnamefont {T.}~\bibnamefont {V.~Trevisan}}, \bibinfo {author} {\bibfnamefont {Y.}~\bibnamefont {Onishi}}, \bibinfo {author} {\bibfnamefont {C.}~\bibnamefont {Hu}}, \bibinfo {author} {\bibfnamefont {T.}~\bibnamefont {Qian}}, \bibinfo {author} {\bibfnamefont {H.-J.}\ \bibnamefont {Tien}}, \bibinfo {author} {\bibfnamefont {S.-W.}\ \bibnamefont {Chen}}, \bibinfo {author} {\bibfnamefont {M.}~\bibnamefont {Huang}}, \bibinfo {author} {\bibfnamefont {D.}~\bibnamefont {B{\'e}rub{\'e}}}, \bibinfo {author} {\bibfnamefont {H.}~\bibnamefont {Li}}, \bibinfo {author} {\bibfnamefont {C.}~\bibnamefont {Tzschaschel}}, \bibinfo {author} {\bibfnamefont {T.}~\bibnamefont {Dinh}}, \bibinfo {author} {\bibfnamefont {Z.}~\bibnamefont {Sun}},
  \bibinfo {author} {\bibfnamefont {S.-C.}\ \bibnamefont {Ho}}, \bibinfo {author} {\bibfnamefont {S.-W.}\ \bibnamefont {Lien}}, \bibinfo {author} {\bibfnamefont {B.}~\bibnamefont {Singh}}, \bibinfo {author} {\bibfnamefont {K.}~\bibnamefont {Watanabe}}, \bibinfo {author} {\bibfnamefont {T.}~\bibnamefont {Taniguchi}}, \bibinfo {author} {\bibfnamefont {D.~C.}\ \bibnamefont {Bell}}, \bibinfo {author} {\bibfnamefont {H.}~\bibnamefont {Lin}}, \bibinfo {author} {\bibfnamefont {T.-R.}\ \bibnamefont {Chang}}, \bibinfo {author} {\bibfnamefont {C.~R.}\ \bibnamefont {Du}}, \bibinfo {author} {\bibfnamefont {A.}~\bibnamefont {Bansil}}, \bibinfo {author} {\bibfnamefont {L.}~\bibnamefont {Fu}}, \bibinfo {author} {\bibfnamefont {N.}~\bibnamefont {Ni}}, \bibinfo {author} {\bibfnamefont {P.~P.}\ \bibnamefont {Orth}}, \bibinfo {author} {\bibfnamefont {Q.}~\bibnamefont {Ma}},\ and\ \bibinfo {author} {\bibfnamefont {S.-Y.}\ \bibnamefont {Xu}},\ }\bibfield  {title} {\textit {\bibinfo {title} {Quantum Metric Nonlinear {{Hall}}
  Effect in a Topological Antiferromagnetic Heterostructure},}\ }\href {https://doi.org/10.1126/science.adf1506} {\bibfield  {journal} {\bibinfo  {journal} {Science}\ }\textbf {\bibinfo {volume} {381}},\ \bibinfo {pages} {181--186} (\bibinfo {year} {2023})}\BibitemShut {NoStop}%
\bibitem [{\citenamefont {Roy}(2014)}]{royBandGeometryFractional2014}%
  \BibitemOpen
  \bibfield  {author} {\bibinfo {author} {\bibfnamefont {R.}~\bibnamefont {Roy}},\ }\bibfield  {title} {\textit {\bibinfo {title} {Band Geometry of Fractional Topological Insulators},}\ }\href {https://doi.org/10.1103/PhysRevB.90.165139} {\bibfield  {journal} {\bibinfo  {journal} {Phys. Rev. B}\ }\textbf {\bibinfo {volume} {90}},\ \bibinfo {pages} {165139} (\bibinfo {year} {2014})}\BibitemShut {NoStop}%
\bibitem [{\citenamefont {Rhim}\ \emph {et~al.}(2020)\citenamefont {Rhim}, \citenamefont {Kim},\ and\ \citenamefont {Yang}}]{Rhim2020a}%
  \BibitemOpen
  \bibfield  {author} {\bibinfo {author} {\bibfnamefont {J.-W.}\ \bibnamefont {Rhim}}, \bibinfo {author} {\bibfnamefont {K.}~\bibnamefont {Kim}},\ and\ \bibinfo {author} {\bibfnamefont {B.-J.}\ \bibnamefont {Yang}},\ }\bibfield  {title} {\textit {\bibinfo {title} {Quantum distance and anomalous Landau levels of flat bands},}\ }\href {https://doi.org/10.1038/s41586-020-2540-1} {\bibfield  {journal} {\bibinfo  {journal} {Nature}\ }\textbf {\bibinfo {volume} {584}},\ \bibinfo {pages} {59--63} (\bibinfo {year} {2020})}\BibitemShut {NoStop}%
\bibitem [{\citenamefont {Peotta}\ and\ \citenamefont {T{\"o}rm{\"a}}(2015)}]{peottaSuperfluidityTopologicallyNontrivial2015}%
  \BibitemOpen
  \bibfield  {author} {\bibinfo {author} {\bibfnamefont {S.}~\bibnamefont {Peotta}}\ and\ \bibinfo {author} {\bibfnamefont {P.}~\bibnamefont {T{\"o}rm{\"a}}},\ }\bibfield  {title} {\textit {\bibinfo {title} {Superfluidity in Topologically Nontrivial Flat Bands},}\ }\href {https://doi.org/10.1038/ncomms9944} {\bibfield  {journal} {\bibinfo  {journal} {Nat. Commun.}\ }\textbf {\bibinfo {volume} {6}},\ \bibinfo {pages} {8944} (\bibinfo {year} {2015})}\BibitemShut {NoStop}%
\bibitem [{\citenamefont {Rossi}(2021)}]{rossiQuantum2021}%
  \BibitemOpen
  \bibfield  {author} {\bibinfo {author} {\bibfnamefont {E.}~\bibnamefont {Rossi}},\ }\bibfield  {title} {\textit {\bibinfo {title} {Quantum Metric and Correlated States in Two-Dimensional Systems},}\ }\href {https://doi.org/10.1016/j.cossms.2021.100952} {\bibfield  {journal} {\bibinfo  {journal} {Curr. Opin. Solid State Mater. Sci.}\ }\textbf {\bibinfo {volume} {25}},\ \bibinfo {pages} {100952} (\bibinfo {year} {2021})}\BibitemShut {NoStop}%
\bibitem [{\citenamefont {Zhang}\ \emph {et~al.}(2011)\citenamefont {Zhang}, \citenamefont {Jung}, \citenamefont {Fiete}, \citenamefont {Niu},\ and\ \citenamefont {MacDonald}}]{zhangSpontaneousQuantumHall2011}%
  \BibitemOpen
  \bibfield  {author} {\bibinfo {author} {\bibfnamefont {F.}~\bibnamefont {Zhang}}, \bibinfo {author} {\bibfnamefont {J.}~\bibnamefont {Jung}}, \bibinfo {author} {\bibfnamefont {G.~A.}\ \bibnamefont {Fiete}}, \bibinfo {author} {\bibfnamefont {Q.}~\bibnamefont {Niu}},\ and\ \bibinfo {author} {\bibfnamefont {A.~H.}\ \bibnamefont {MacDonald}},\ }\bibfield  {title} {\textit {\bibinfo {title} {Spontaneous {{Quantum Hall States}} in {{Chirally Stacked Few-Layer Graphene Systems}}},}\ }\href {https://doi.org/10.1103/PhysRevLett.106.156801} {\bibfield  {journal} {\bibinfo  {journal} {Phys. Rev. Lett.}\ }\textbf {\bibinfo {volume} {106}},\ \bibinfo {pages} {156801} (\bibinfo {year} {2011})}\BibitemShut {NoStop}%
\bibitem [{\citenamefont {Ahn}\ \emph {et~al.}(2022)\citenamefont {Ahn}, \citenamefont {Guo}, \citenamefont {Nagaosa},\ and\ \citenamefont {Vishwanath}}]{ahnRiemannianGeometryResonant2022}%
  \BibitemOpen
  \bibfield  {author} {\bibinfo {author} {\bibfnamefont {J.}~\bibnamefont {Ahn}}, \bibinfo {author} {\bibfnamefont {G.-Y.}\ \bibnamefont {Guo}}, \bibinfo {author} {\bibfnamefont {N.}~\bibnamefont {Nagaosa}},\ and\ \bibinfo {author} {\bibfnamefont {A.}~\bibnamefont {Vishwanath}},\ }\bibfield  {title} {\textit {\bibinfo {title} {Riemannian Geometry of Resonant Optical Responses},}\ }\href {https://doi.org/10.1038/s41567-021-01465-z} {\bibfield  {journal} {\bibinfo  {journal} {Nat. Phys.}\ }\textbf {\bibinfo {volume} {18}},\ \bibinfo {pages} {290--295} (\bibinfo {year} {2022})}\BibitemShut {NoStop}%
\bibitem [{\citenamefont {Hasan}\ and\ \citenamefont {Kane}(2010)}]{hasanColloquium2010}%
  \BibitemOpen
  \bibfield  {author} {\bibinfo {author} {\bibfnamefont {M.~Z.}\ \bibnamefont {Hasan}}\ and\ \bibinfo {author} {\bibfnamefont {C.~L.}\ \bibnamefont {Kane}},\ }\bibfield  {title} {\textit {\bibinfo {title} {{\emph{Colloquium}} : {{Topological}} Insulators},}\ }\href {https://doi.org/10.1103/RevModPhys.82.3045} {\bibfield  {journal} {\bibinfo  {journal} {Rev. Mod. Phys.}\ }\textbf {\bibinfo {volume} {82}},\ \bibinfo {pages} {3045--3067} (\bibinfo {year} {2010})}\BibitemShut {NoStop}%
\bibitem [{\citenamefont {Bl{\aa}sj{\"o}}(2005)}]{blasjoIsoperimetric2005a}%
  \BibitemOpen
  \bibfield  {author} {\bibinfo {author} {\bibfnamefont {V.}~\bibnamefont {Bl{\aa}sj{\"o}}},\ }\bibfield  {title} {\textit {\bibinfo {title} {The {{Isoperimetric Problem}}},}\ }\href {https://doi.org/10.2307/30037526} {\bibfield  {journal} {\bibinfo  {journal} {Amer. Math. Monthly}\ }\textbf {\bibinfo {volume} {112}},\ \bibinfo {pages} {526--566} (\bibinfo {year} {2005})}\BibitemShut {NoStop}%
\bibitem [{\citenamefont {Osserman}(1978)}]{ossermanIsoperimetricInequality1978}%
  \BibitemOpen
  \bibfield  {author} {\bibinfo {author} {\bibfnamefont {R.}~\bibnamefont {Osserman}},\ }\bibfield  {title} {\textit {\bibinfo {title} {The Isoperimetric Inequality},}\ }\href {https://doi.org/10.1090/S0002-9904-1978-14553-4} {\bibfield  {journal} {\bibinfo  {journal} {Bull. Amer. Math. Soc.}\ }\textbf {\bibinfo {volume} {84}},\ \bibinfo {pages} {1182--1238} (\bibinfo {year} {1978})}\BibitemShut {NoStop}%
\bibitem [{\citenamefont {Gelfand}\ \emph {et~al.}(2000)\citenamefont {Gelfand}, \citenamefont {Fomin},\ and\ \citenamefont {Silverman}}]{gelfandCalculusVariations2000}%
  \BibitemOpen
  \bibfield  {author} {\bibinfo {author} {\bibfnamefont {I.~M.}\ \bibnamefont {Gelfand}}, \bibinfo {author} {\bibfnamefont {S.~V.}\ \bibnamefont {Fomin}},\ and\ \bibinfo {author} {\bibfnamefont {R.~A.}\ \bibnamefont {Silverman}},\ }\href@noop {} {\emph {\bibinfo {title} {Calculus of Variations}}},\ \bibinfo {edition} {unabridged repr}\ ed.\ (\bibinfo  {publisher} {Dover publ},\ \bibinfo {address} {New York},\ \bibinfo {year} {2000})\BibitemShut {NoStop}%
\bibitem [{\citenamefont {Cheng}(2013)}]{chengQuantum2010}%
  \BibitemOpen
  \bibfield  {author} {\bibinfo {author} {\bibfnamefont {R.}~\bibnamefont {Cheng}},\ }\textit {\bibinfo {title} {Quantum Geometric Tensor (Fubini-Study Metric) in Simple Quantum System: A pedagogical Introduction},}\ \Eprint {https://arxiv.org/abs/1012.1337} {arXiv:1012.1337}\BibitemShut {NoStop}%
\bibitem [{\citenamefont {Onishi}\ and\ \citenamefont {Fu}(2024)}]{onishiFundamentalBoundTopological2024}%
  \BibitemOpen
  \bibfield  {author} {\bibinfo {author} {\bibfnamefont {Y.}~\bibnamefont {Onishi}}\ and\ \bibinfo {author} {\bibfnamefont {L.}~\bibnamefont {Fu}},\ }\bibfield  {title} {\textit {\bibinfo {title} {Fundamental {{Bound}} on {{Topological Gap}}},}\ }\href {https://doi.org/10.1103/PhysRevX.14.011052} {\bibfield  {journal} {\bibinfo  {journal} {Phys. Rev. X}\ }\textbf {\bibinfo {volume} {14}},\ \bibinfo {pages} {011052} (\bibinfo {year} {2024})}\BibitemShut {NoStop}%
\bibitem [{\citenamefont {Yu}\ \emph {et~al.}(2024)\citenamefont {Yu}, \citenamefont {Ciccarino}, \citenamefont {Bianco}, \citenamefont {Errea}, \citenamefont {Narang},\ and\ \citenamefont {Bernevig}}]{yuNontrivialQuantumGeometry2024}%
  \BibitemOpen
  \bibfield  {author} {\bibinfo {author} {\bibfnamefont {J.}~\bibnamefont {Yu}}, \bibinfo {author} {\bibfnamefont {C.~J.}\ \bibnamefont {Ciccarino}}, \bibinfo {author} {\bibfnamefont {R.}~\bibnamefont {Bianco}}, \bibinfo {author} {\bibfnamefont {I.}~\bibnamefont {Errea}}, \bibinfo {author} {\bibfnamefont {P.}~\bibnamefont {Narang}},\ and\ \bibinfo {author} {\bibfnamefont {B.~A.}\ \bibnamefont {Bernevig}},\ }\bibfield  {title} {\textit {\bibinfo {title} {Non-Trivial Quantum Geometry and the Strength of Electron--Phonon Coupling},}\ }\href {https://doi.org/10.1038/s41567-024-02486-0} {\bibfield  {journal} {\bibinfo  {journal} {Nat. Phys.}\ }\textbf {\bibinfo {volume} {20}},\ \bibinfo {pages} {1262--1268} (\bibinfo {year} {2024})}\BibitemShut {NoStop}%
\bibitem [{\citenamefont {Nakahara}(2005)}]{nakaharaGeometryTopologyPhysics2005}%
  \BibitemOpen
  \bibfield  {author} {\bibinfo {author} {\bibfnamefont {M.}~\bibnamefont {Nakahara}},\ }\href@noop {} {\emph {\bibinfo {title} {Geometry, Topology, and Physics}}},\ \bibinfo {edition} {2nd}\ ed.,\ Graduate Student Series in Physics\ (\bibinfo  {publisher} {Inst. of Physics Publishing},\ \bibinfo {address} {Bristol},\ \bibinfo {year} {2005})\BibitemShut {NoStop}%
\bibitem [{\citenamefont {Xiao}\ \emph {et~al.}(2010)\citenamefont {Xiao}, \citenamefont {Chang},\ and\ \citenamefont {Niu}}]{xiaoBerryPhaseEffects2010}%
  \BibitemOpen
  \bibfield  {author} {\bibinfo {author} {\bibfnamefont {D.}~\bibnamefont {Xiao}}, \bibinfo {author} {\bibfnamefont {M.-C.}\ \bibnamefont {Chang}},\ and\ \bibinfo {author} {\bibfnamefont {Q.}~\bibnamefont {Niu}},\ }\bibfield  {title} {\textit {\bibinfo {title} {Berry Phase Effects on Electronic Properties},}\ }\href {https://doi.org/10.1103/RevModPhys.82.1959} {\bibfield  {journal} {\bibinfo  {journal} {Rev. Mod. Phys.}\ }\textbf {\bibinfo {volume} {82}},\ \bibinfo {pages} {1959--2007} (\bibinfo {year} {2010})}\BibitemShut {NoStop}%
\bibitem [{\citenamefont {Marzari}\ and\ \citenamefont {Vanderbilt}(1997)}]{marzariMaximallyLocalizedGeneralized1997}%
  \BibitemOpen
  \bibfield  {author} {\bibinfo {author} {\bibfnamefont {N.}~\bibnamefont {Marzari}}\ and\ \bibinfo {author} {\bibfnamefont {D.}~\bibnamefont {Vanderbilt}},\ }\bibfield  {title} {\textit {\bibinfo {title} {Maximally Localized Generalized {{Wannier}} Functions for Composite Energy Bands},}\ }\href {https://doi.org/10.1103/PhysRevB.56.12847} {\bibfield  {journal} {\bibinfo  {journal} {Phys. Rev. B}\ }\textbf {\bibinfo {volume} {56}},\ \bibinfo {pages} {12847--12865} (\bibinfo {year} {1997})}\BibitemShut {NoStop}%
\bibitem [{\citenamefont {{King-Smith}}\ and\ \citenamefont {Vanderbilt}(1993)}]{king-smithTheoryPolarizationCrystalline1993}%
  \BibitemOpen
  \bibfield  {author} {\bibinfo {author} {\bibfnamefont {R.~D.}\ \bibnamefont {{King-Smith}}}\ and\ \bibinfo {author} {\bibfnamefont {D.}~\bibnamefont {Vanderbilt}},\ }\bibfield  {title} {\textit {\bibinfo {title} {Theory of Polarization of Crystalline Solids},}\ }\href {https://doi.org/10.1103/PhysRevB.47.1651} {\bibfield  {journal} {\bibinfo  {journal} {Phys. Rev. B}\ }\textbf {\bibinfo {volume} {47}},\ \bibinfo {pages} {1651--1654} (\bibinfo {year} {1993})}\BibitemShut {NoStop}%
\bibitem [{\citenamefont {Deffner}\ and\ \citenamefont {Campbell}(2017)}]{deffnerQuantumSpeedLimits2017}%
  \BibitemOpen
  \bibfield  {author} {\bibinfo {author} {\bibfnamefont {S.}~\bibnamefont {Deffner}}\ and\ \bibinfo {author} {\bibfnamefont {S.}~\bibnamefont {Campbell}},\ }\bibfield  {title} {\textit {\bibinfo {title} {Quantum Speed Limits: From {{Heisenberg}}'s Uncertainty Principle to Optimal Quantum Control},}\ }\href {https://doi.org/10.1088/1751-8121/aa86c6} {\bibfield  {journal} {\bibinfo  {journal} {J. Phys. A}\ }\textbf {\bibinfo {volume} {50}},\ \bibinfo {pages} {453001} (\bibinfo {year} {2017})}\BibitemShut {NoStop}%
\bibitem [{\citenamefont {Bohm}\ and\ \citenamefont {Sato}(2005)}]{bohmRelativistic2005}%
  \BibitemOpen
  \bibfield  {author} {\bibinfo {author} {\bibfnamefont {A.~R.}\ \bibnamefont {Bohm}}\ and\ \bibinfo {author} {\bibfnamefont {Y.}~\bibnamefont {Sato}},\ }\bibfield  {title} {\textit {\bibinfo {title} {Relativistic Resonances: {{Their}} Masses, Widths, Lifetimes, Superposition, and Causal Evolution},}\ }\href {https://doi.org/10.1103/PhysRevD.71.085018} {\bibfield  {journal} {\bibinfo  {journal} {Phys. Rev. D}\ }\textbf {\bibinfo {volume} {71}},\ \bibinfo {pages} {085018} (\bibinfo {year} {2005})}\BibitemShut {NoStop}%
\bibitem [{\citenamefont {Mandelstam}\ and\ \citenamefont {Tamm}(1945)}]{mandelstamUncertainty1991}%
  \BibitemOpen
  \bibfield  {author} {\bibinfo {author} {\bibfnamefont {L.}~\bibnamefont {Mandelstam}}\ and\ \bibinfo {author} {\bibfnamefont {I.}~\bibnamefont {Tamm}},\ }\bibfield  {title} {\textit {\bibinfo {title} {The {{Uncertainty Relation Between Energy}} and {{Time}} in {{Non-relativistic Quantum Mechanics}}},}\ }\href@noop {} {\bibfield  {journal} {\bibinfo  {journal} {J. Phys. USSR}\ }\textbf {\bibinfo {volume} {9}},\ \bibinfo {pages} {249--254} (\bibinfo {year} {1945})}\BibitemShut {NoStop}%
\bibitem [{\citenamefont {Margolus}\ and\ \citenamefont {Levitin}(1998)}]{margolusmaximum1998}%
  \BibitemOpen
  \bibfield  {author} {\bibinfo {author} {\bibfnamefont {N.}~\bibnamefont {Margolus}}\ and\ \bibinfo {author} {\bibfnamefont {L.~B.}\ \bibnamefont {Levitin}},\ }\bibfield  {title} {\textit {\bibinfo {title} {The Maximum Speed of Dynamical Evolution},}\ }\href {https://doi.org/10.1016/S0167-2789(98)00054-2} {\bibfield  {journal} {\bibinfo  {journal} {Physica D}\ }\textbf {\bibinfo {volume} {120}},\ \bibinfo {pages} {188--195} (\bibinfo {year} {1998})}\BibitemShut {NoStop}%
\bibitem [{\citenamefont {Anandan}\ and\ \citenamefont {Aharonov}(1990)}]{anandanGeometryQuantumEvolution1990}%
  \BibitemOpen
  \bibfield  {author} {\bibinfo {author} {\bibfnamefont {J.}~\bibnamefont {Anandan}}\ and\ \bibinfo {author} {\bibfnamefont {Y.}~\bibnamefont {Aharonov}},\ }\bibfield  {title} {\textit {\bibinfo {title} {Geometry of Quantum Evolution},}\ }\href {https://doi.org/10.1103/PhysRevLett.65.1697} {\bibfield  {journal} {\bibinfo  {journal} {Phys. Rev. Lett.}\ }\textbf {\bibinfo {volume} {65}},\ \bibinfo {pages} {1697--1700} (\bibinfo {year} {1990})}\BibitemShut {NoStop}%
\bibitem [{\citenamefont {Allen}\ and\ \citenamefont {Dynes}(1975)}]{allenTransitionTemperatureStrongcoupled1975}%
  \BibitemOpen
  \bibfield  {author} {\bibinfo {author} {\bibfnamefont {P.~B.}\ \bibnamefont {Allen}}\ and\ \bibinfo {author} {\bibfnamefont {R.~C.}\ \bibnamefont {Dynes}},\ }\bibfield  {title} {\textit {\bibinfo {title} {Transition Temperature of Strong-Coupled Superconductors Reanalyzed},}\ }\href {https://doi.org/10.1103/PhysRevB.12.905} {\bibfield  {journal} {\bibinfo  {journal} {Phys. Rev. B}\ }\textbf {\bibinfo {volume} {12}},\ \bibinfo {pages} {905--922} (\bibinfo {year} {1975})}\BibitemShut {NoStop}%
\bibitem [{\citenamefont {Nambu}(1960)}]{nambuQuasiParticlesGaugeInvariance1960}%
  \BibitemOpen
  \bibfield  {author} {\bibinfo {author} {\bibfnamefont {Y.}~\bibnamefont {Nambu}},\ }\bibfield  {title} {\textit {\bibinfo {title} {Quasi-{{Particles}} and {{Gauge Invariance}} in the {{Theory}} of {{Superconductivity}}},}\ }\href {https://doi.org/10.1103/PhysRev.117.648} {\bibfield  {journal} {\bibinfo  {journal} {Phys. Rev.}\ }\textbf {\bibinfo {volume} {117}},\ \bibinfo {pages} {648--663} (\bibinfo {year} {1960})}\BibitemShut {NoStop}%
\bibitem [{\citenamefont {Tovmasyan}\ \emph {et~al.}(2016)\citenamefont {Tovmasyan}, \citenamefont {Peotta}, \citenamefont {T{\"o}rm{\"a}},\ and\ \citenamefont {Huber}}]{tovmasyanEffectiveTheoryEmergent2016}%
  \BibitemOpen
  \bibfield  {author} {\bibinfo {author} {\bibfnamefont {M.}~\bibnamefont {Tovmasyan}}, \bibinfo {author} {\bibfnamefont {S.}~\bibnamefont {Peotta}}, \bibinfo {author} {\bibfnamefont {P.}~\bibnamefont {T{\"o}rm{\"a}}},\ and\ \bibinfo {author} {\bibfnamefont {S.~D.}\ \bibnamefont {Huber}},\ }\bibfield  {title} {\textit {\bibinfo {title} {Effective Theory and Emergent {{SU}} ( 2 ) Symmetry in the Flat Bands of Attractive {{Hubbard}} Models},}\ }\href {https://doi.org/10.1103/PhysRevB.94.245149} {\bibfield  {journal} {\bibinfo  {journal} {Phys. Rev. B}\ }\textbf {\bibinfo {volume} {94}},\ \bibinfo {pages} {245149} (\bibinfo {year} {2016})}\BibitemShut {NoStop}%
\bibitem [{\citenamefont {Su}\ \emph {et~al.}(1979)\citenamefont {Su}, \citenamefont {Schrieffer},\ and\ \citenamefont {Heeger}}]{suSolitonsPolyacetylene1979}%
  \BibitemOpen
  \bibfield  {author} {\bibinfo {author} {\bibfnamefont {W.~P.}\ \bibnamefont {Su}}, \bibinfo {author} {\bibfnamefont {J.~R.}\ \bibnamefont {Schrieffer}},\ and\ \bibinfo {author} {\bibfnamefont {A.~J.}\ \bibnamefont {Heeger}},\ }\bibfield  {title} {\textit {\bibinfo {title} {Solitons in {{Polyacetylene}}},}\ }\href {https://doi.org/10.1103/PhysRevLett.42.1698} {\bibfield  {journal} {\bibinfo  {journal} {Phys. Rev. Lett.}\ }\textbf {\bibinfo {volume} {42}},\ \bibinfo {pages} {1698--1701} (\bibinfo {year} {1979})}\BibitemShut {NoStop}%
\bibitem [{\citenamefont {Creutz}(1999)}]{creutzEndStatesLadder1999}%
  \BibitemOpen
  \bibfield  {author} {\bibinfo {author} {\bibfnamefont {M.}~\bibnamefont {Creutz}},\ }\bibfield  {title} {\textit {\bibinfo {title} {End {{States}}, {{Ladder Compounds}}, and {{Domain-Wall Fermions}}},}\ }\href {https://doi.org/10.1103/PhysRevLett.83.2636} {\bibfield  {journal} {\bibinfo  {journal} {Phys. Rev. Lett.}\ }\textbf {\bibinfo {volume} {83}},\ \bibinfo {pages} {2636--2639} (\bibinfo {year} {1999})}\BibitemShut {NoStop}%
\bibitem [{\citenamefont {Mondaini}\ \emph {et~al.}(2018)\citenamefont {Mondaini}, \citenamefont {Batrouni},\ and\ \citenamefont {Gr{\'e}maud}}]{mondainiPairingSuperconductivityFlat2018}%
  \BibitemOpen
  \bibfield  {author} {\bibinfo {author} {\bibfnamefont {R.}~\bibnamefont {Mondaini}}, \bibinfo {author} {\bibfnamefont {G.~G.}\ \bibnamefont {Batrouni}},\ and\ \bibinfo {author} {\bibfnamefont {B.}~\bibnamefont {Gr{\'e}maud}},\ }\bibfield  {title} {\textit {\bibinfo {title} {Pairing and Superconductivity in the Flat Band: {{Creutz}} Lattice},}\ }\href {https://doi.org/10.1103/PhysRevB.98.155142} {\bibfield  {journal} {\bibinfo  {journal} {Phys. Rev. B}\ }\textbf {\bibinfo {volume} {98}},\ \bibinfo {pages} {155142} (\bibinfo {year} {2018})}\BibitemShut {NoStop}%
\end{thebibliography}%
\end{document}